\documentclass[10pt, conference, letterpaper]{IEEEtran}

\IEEEoverridecommandlockouts                              

\usepackage{amsmath}
\usepackage{graphics}
\usepackage{graphicx}
\usepackage{epsfig,amssymb}
\usepackage{epstopdf}
\usepackage{amsmath}
\usepackage{times}
\usepackage{mathrsfs}
\usepackage{array}
\usepackage{amsmath, latexsym, amsfonts, amssymb}
\usepackage[mathscr]{eucal}
\usepackage{array, tabularx}
\usepackage[table]{xcolor}
\usepackage{multirow}
\usepackage{epsfig}
\usepackage{cite}
\usepackage{tikz}
\usepackage{url}
\usepackage{mathtools}
\usepackage{psfragx}
\usepackage{xcolor}
\usepackage{bbm}
\usepackage{lipsum}
\usepackage{multicol}
\usepackage{amssymb}
\usepackage[linesnumbered,ruled]{algorithm2e}
\usepackage{color,soul}

\title{\LARGE \bf
Towards Comfortable Cycling: A Practical Approach to Monitor the Conditions in Cycling Paths
}

\author{\IEEEauthorblockN{Nipun Wijerathne, 
		Sanjana Kadaba Viswanath,
		Marakkalage Sumudu Hasala,
		Victoria Beltran,				
		Chau Yuen,
		Hock Lim	
	}
	\IEEEauthorblockA{
		Engineering Product Development, Singapore University of Technology and Design, Singapore		
		\\
		Email: \{gallage\_wijerathne,sanjana,maria\_beltran,yuenchau,hockbeng\_lim\}@sutd.edu.sg,   
	    marakkalage@mymail.sutd.edu.sg,	    
	}
	
		\thanks{This work was supported in part by the International Design Center, Municipal Services Office (MSO) and in part by the National Science Foundation of China (NSFC) (61750110529).}
	
	}

 \vspace{-9mm}

\begin{document}

\IEEEoverridecommandlockouts
\IEEEpubid{\makebox[\columnwidth]{978-1-4673-9944-9/18/$31.00 \copyright~2018~IEEE }
	\hspace{\columnsep}\makebox[\columnwidth]{ }}

\maketitle
\thispagestyle{empty}
\pagestyle{empty}

\begin{abstract}
This is a no brainer. Using bicycles to commute is the most sustainable form of transport, is the least expensive to use and are pollution-free. Towns and cities have to be made bicycle-friendly to encourage their wide usage. Therefore, cycling paths should be more convenient, comfortable, and safe to ride. This paper investigates a smartphone application, which passively monitors the road conditions during cyclists ride. To overcome the problems of monitoring roads, we present novel algorithms that sense the rough cycling paths and locate road bumps. Each event is detected in real time to improve the user friendliness of the application. Cyclists may keep their smartphones at any random orientation and placement. Moreover, different smartphones sense the same incident dissimilarly and hence report discrepant sensor values. We further address the aforementioned difficulties that limit such crowd-sourcing application. We evaluate our sensing application on cycling paths in Singapore, and show that it can successfully detect such bad road conditions.

\end{abstract}

\section{Introduction}\label{ID}


Transportation has become an integral part of the daily lives of people. Now, most of the countries' intention in which to go car-lite nation has now gained much attention where people are progressively moving towards the public transportation, use of personal mobility devices (\textit{PMD}), and cycling. Therefore, to encourage adoption of \textit{PMD} and cycling as an essential mode of transportation, more convenient, comfortable, and safer infrastructures should be developed and expanded.

Dangerous road surface conditions with bumpy roads, potholes and rough roads are the major distractions for safe, and comfortable cycling. Therefore, it is a responsibility of the municipal councils to monitor the bad road conditions and take appropriate actions to fix them. Contemporary, monitoring such conditions is challenging as it is highly labour intensive, time-consuming and expensive. Therefore, prior work in this area has primarily focused on automating such a process with the help of sensing devices such as accelerometer, gyroscope, magnetometer, and \textit{GPS}. \cite{pothole, recursive, crowdsourcing, cond} use customised devices with one or more of above sensors to monitor the road conditions. However, building and manoeuvring such customised infrastructures are expensive and arduous. Therefore, most of the recent works revolve around smartphones and crowd-sourcing approaches \cite{time, rich, explore, speed,ran_ieee_sensors2017,Yassin_ieee_tutorials2016}.

\begin{figure}[ht]
	\centering
	\includegraphics[width=0.35\textwidth]{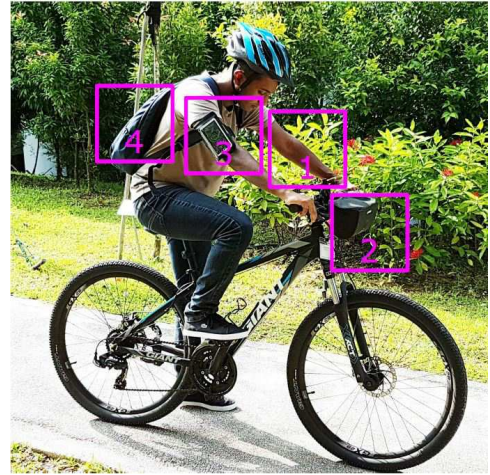}
	\caption{A cyclist with different smart phone placements. (1) mounted. (2) front bag. (3) hand pouch (4) backpack.}	 
	\label{fig1}
	\vspace{-0.5cm}
\end{figure}

Modern smartphones now come up with many rich sensing and communication capabilities that include at least an accelerometer, \textit{GPS}, and internet access. Notice that various smartphones employ different types of motion sensors. For instance, several smartphones may only have the accelerometer sensor whereas remaining will have one or more sophisticated motion sensors like gyroscope and magnetometer. Moreover, the number of smartphone subscribers and mobile internet users are growing exponentially and hence a smartphone-based approach is truly realistic for our purpose. Therefore, we developed a smartphone application for cyclists that allow us to collect data passively about the roads that cyclists ride on. Cyclists then only need to carry the smartphone whenever they cycle, and our application keeps the track of road conditions.

One of the major challenges is that cyclists could carry or keep their mobile phones in any location of the bicycle. However, based on a survey that we conducted, most common places where cyclists keep their phones are shown in  Fig. \ref{fig1}. The smartphone could be kept at one of place from $ 1) $ mounted in a mounter $ 2) $ in a front bag $ 3) $ in a hand pouch $ 4) $ in a backpack. It should be noted that aforementioned places are only the general places and hence algorithms that monitor the road conditions should not be restricted to them. Moreover, a smartphone could be kept at any arbitrary orientation by a cyclist. Therefore, algorithms should be capable of handling any random placement and orientation. More detailed discussion on this matter of contention can be found in Section \ref{SO}.

\begin{figure*}[t]
	\fontsize{8pt}{8pt}\selectfont
	\begin{tabular}{cc}		
		\includegraphics[width=0.45\textwidth]{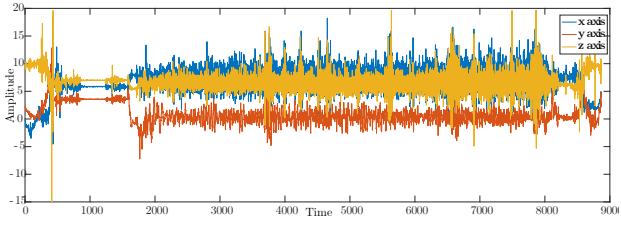}&   
		\includegraphics[width=0.45\textwidth]{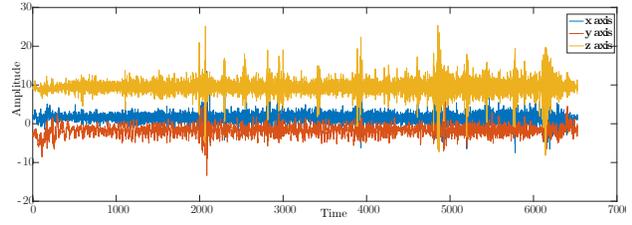}    
	\end{tabular}
	\caption{Left figure is 3-axis accelerometer signal when smart phone is mounted and right figure is the same signal when smart phone is placed on the front bag.}
	\label{fig2}
	\vspace{-0.5cm}
\end{figure*}

In this paper, we focus on monitoring the road conditions by sensing road roughness levels and locating road bumps. The smartphone application locates each event via \textit{GPS} and sends back to the server with the time-stamp and event intensity at the end of each cycling trip. Therefore, we could aggregate events from many cyclists and remove any isolated events, which could be possible false alarms. Moreover, we minimise the mobile data usage and battery draining by detecting each event in real time and hence the user friendliness of the application is improved. The contributions of our work are two-fold:
\begin{itemize}
	\item We propose novel algorithms that sense rough road conditions and locate road bumps, regardless of the placement or orientation of the smartphone.
	\item We have implemented the aforementioned algorithms in a \textit{Android} based system that could distribute among public to gather road information (crowd-sourcing).
\end{itemize}
The rest of this paper is organised as follows. Next Section discusses related works in detail. Section \ref{SO} presents more about smartphone orientation and placements, which will be followed by the details of the low pass filter and wavelet transformation. Section \ref{RCM} presents algorithms to sense rough road conditions and locate road bumps.
\section{Related work}\label{RW}
As noted in Section \ref{ID}, customised infrastructure and smartphone-based applications can be commonly found in the literature. Both approaches use \textit{GPS} to sense the location, and the algorithms are based on the information provided by at least an accelerometer in which sometimes coupled with gyroscope and magnetometer. Most of the previous works have done in the context of road bump detection and a limited amount of work in the areas of rough road sensing.

\subsection{rough road sensing}
Rough road sensing algorithms, which are based on signal processing approaches can be found commonly in literature. \cite{cond} uses power spectral density to estimate the standard deviation (\textit{SD}) of the $ Zaxis $ accelerometer signal. Then a roughness index is derived from \textit{SD}. Similarly, \cite{crowdsourcing} introduces \textit{i}-\textit{GMM} algorithm that is based on Gaussian mixture models to determine the rough road conditions. Therefore, \cite{crowdsourcing} estimates the weight, mean and variance of each Gaussian model and hence determines the roughness from the accelerometer signal if it deviates significantly from the pre-estimated parameters. Moreover, \cite{cond} and \cite{crowdsourcing} use a customised device to collect accelerometer and \textit{GPS} data and the device is mounted at a known orientation. 

\subsection{bump detection}
Most of the previous bump detection algorithms use only the accelerometer data. Therefore, except \cite{rich} that uses a virtual reorientation method, all other algorithms require keeping the smartphone at a fixed known orientation when acquiring the accelerometer data. In contrast, the virtual reorientation method use by \cite{rich} reorientates any unknown orientation. However, the reorientation algorithm searches for stationary and large acceleration/deceleration events of the smartphone, which limits the applicability. Most of the previous algorithms can be categorised as follows.
\subsubsection{Z-threshold algorithms}Algorithm classifies any event as a bump if the $ Zaxis $ (see Section \ref{SO} for more details) value exceeds a certain threshold. The threshold is determined empirically, and relatively the same versions of the algorithm can be found in\cite{rich, pothole}.
\subsubsection{Z-difference algorithms}Algorithm determines a bump if the difference between two successive $ Zaxis $ readings exceed a certain threshold. Moreover, \cite{time} claims that it is as an improved version of the \textit{Z-threshold} algorithm.
\subsubsection{Algorithms based on data mining} In \cite{classification}, first trains a suitable classification model using training data with carefully acquired ground truth. Then the estimated model is used to detect bumps in real time. 
\section{System Overview}\label{SO}

Smartphone sensor framework uses $ 3 $-$ axis $ orthogonal coordinate system, which is related to the device default orientation to express its data values. When the device is held in vertical position (\textit{i.e.} default orientation), $ Y axis $ is pointed upwards, $ X axis  $ is pointed right (when it measures in the clockwise direction from $ Y axis $) and, $ Z axis  $ is pointed toward the outside of screen face. Throughout this paper, accelerometer readings along the 3 axes are denoted by $a_{X},a_{Y},$ and $ a_{Z} $. Conjointly, we set the accelerometer sampling frequency to $ 50Hz $ for the rest of our work presented in this paper.

\begin{figure}[ht]
	\centering
	\includegraphics[width=0.45\textwidth]{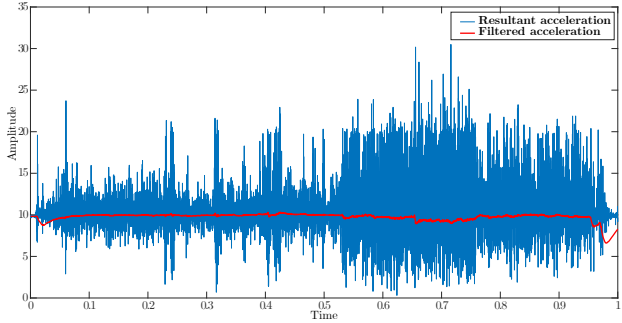}
	\caption{Filtered accelerometer signal and raw accelerometer signal.}	 
	\label{fig4}
\end{figure}

It is important to understand that the smartphone could be held in any arbitrary orientation aside from the default orientation. Moreover, the axes of above coordinate system are not swapped when the device's screen orientation changes. Therefore, the accelerometer readings are no longer valid in the aforementioned directions. It should be noted that our accelerometer measures the force exert on the sensor but not the physical acceleration experienced by the mobile. For instance, the acceleration reported due to the gravitational force is positive when the mobile is held in its default orientation but not negative although the $ Y axis $ is pointed upwards.

While the smartphone is stationary, the resultant acceleration $ a_{R} $ can be calculated according to the Eqn. \ref{eqn1} is reported around $ 1g $ $(9.8 \text{ } ms^{-2})$ regardless of the device's arbitrary orientation. 
\begin{equation} \label{eqn1}
a_{R} = \sqrt{a_{X}^{2}+a_{Y}^{2}+a_{Z}^{2}}
\end{equation}
However, the accelerometer can report a value less than or greater than $ 1g $ in the time of acceleration depending on its orientation. For instance, as noted in Section \ref{ID}, cyclists may place their smartphones in a hand pouch or mounted during their trips. As shown in Fig. \ref{fig2}, two accelerometers report discrepant readings as a result of the dissimilarity between orientations. Therefore, it is foremost important to tackle this unknown orientation problem, which is a typical scenario for smartphones. Most of the previous work \cite{pothole,time} use a fixed known orientation and \cite{rich} uses a framework based on Euler angles to reorient the disoriented accelerometer. In the following section, we leverage a much simpler and pragmatic approach to determine the acceleration component perpendicular to the ground plane \textit{i.e.} $a_{Rg}$. In addition to the gravity, $a_{Rg}$ consists of moderated abrupt changes that could exist due to unevenness of the road.  

\subsection{low-pass filter}\label{RRD}
We exploit a simple first order infinite impulse response \textit{IIR} filter to determine the $a_{Rg}$. The intuition behind is, by applying a low-pass filter, we isolate the low-frequency features and trends \textit{i.e.} the gravity and moderated abrupt changes \cite{gravity}. Moreover, the accelerometer suffers from the drift ($1/f$ noise) at low frequencies and Gaussian noise at high frequencies. Therefore, we apply the same filter to remove the unwanted noise component $a_{W}$ of the signal. In this section, we present the implementation, impulse repose analysis, and the stability analysis of the proposed filter.

The recurrence relation of the \textit{IIR} filter for each axis $ X,Y $ and, $ Z$ can be given as difference equations as follows,
\begin{equation} \label{eqn2}
a_{Xg}[n] = \alpha * a_{Xg}[n-1]+(1-\alpha)* a_{X}[n],
\end{equation}
\begin{equation} \label{eqn3}
a_{Yg}[n] = \alpha * a_{Yg}[n-1]+(1-\alpha)*  a_{Y}[n],
\end{equation}
\begin{equation} \label{eqn4}
a_{Zg}[n] = \alpha * a_{Zg}[n-1]+(1-\alpha)*  a_{Z}[n], 
\end{equation}
where $a_{Xg},a_{Yg},a_{Zg}$ are the filtered signals of inputs $a_{X},a_{Y},a_{Z}$ and $ \alpha $ is the smoothing factor.  We set $ \alpha $ to 0.992, unless otherwise noted in this paper.  $a_{Rg}$ is calculated similar to Eqn. \ref{eqn1} can be given as,
\begin{equation} \label{eqn5}
a_{Rg} = \sqrt{a_{Xg}^{2}+a_{Yg}^{2}+a_{Zg}^{2}}.
\end{equation}
Fig. \ref{fig4} depicts the input $a_{R}$ and the filtered output $a_{Rg}$. Notice that $a_{R}$ consists of linear acceleration $a_{L}$ aside from  the $a_{Rg}$ and $a_{W}$ components. It is not hard to see that our method successfully filtered out the $a_{Rg}$ component that varies around $ 1g $.

 \subsection{Wavelet Transformation} \label{WT}
We divide $ a_{Rg} $ into $ N $ points segments (denoted as  $ \hat a_{Rg}$) and transformed each $ \hat a_{Rg}$ into wavelet domain in order to facilitate rough road sensing and bump detection algorithms that will be discussed in later sections. We set $ N=32 $ carefully, by considering the calculation efficiency of the wavelet transformation. Wavelet coefficients of $ \hat a_{Rg}$ are computed using series of dilation and translation of the mother wavelet\cite{intro_wave}\cite{waver}, \textit{i.e.},
\begin{equation}\label{eqn9}
W\hat a_{Rg}(n) = \sum_{j}\sum_{k}d_{j,k}\psi_{j,k}(n), 
\end{equation}
where $\psi_{j,k}(n)$ is the scaled and dilated mother wavelet function, $d_{j,k}$ is the wavelet coefficient that represents how much translated and dialed mother wavelet describes the given $\hat a_{Rg}(n)$, and $ j,k \in \mathbb{Z} $. To this end, we have  
\begin{equation}\label{eqn10}
d_{j,k}= \sum_{n}\hat a_{Rg}(n)\psi_{j,k}(n)= <\hat a_{Rg}, \psi_{j,k}>,
\end{equation}
since wavelets are orthogonal to each other. More detailed explanation on wavelets can be found in  \cite{waver}. In this paper, we use \textit{Haar} basis function as the mother wavelet due to its own discontinuous nature, \textit{i.e.},
\begin{equation}\label{eqn11}
\psi_{j,k}(x) = \begin{cases}
\enspace 1  & \text{if } x \in [0,1/2) \\
-1  & \text{if } x \in [1/2,1) \\
\enspace 0  & \text{otherwise } 
\end{cases}.
\end{equation}
Therefore, we represent each $ \hat a_{Rg}$ in terms of  \textit{Haar} basis function, and those wavelet coefficients $\{d_{j,k} \}_{j,k \in \mathbb{Z}}$ are used for further calculations. In this paper, we calculate $\{d_{j,k} \}_{j,k \in \mathbb{Z}}$ according to  \cite{nondy}. To this end, discrete \textit{Haar} wavelet coefficients can be calculated as 
\begin{equation}\label{eqn12}
\bf W\hat a_{Rg}=B^{-1}\hat a_{Rg},
\end{equation}
where $ \bf B $ is the wavelet basis matrix in \cite{nondy}. Although the time complexity of above calculation can be given as $\mathcal{O}(n^2)$, the efficiency of the calculation is still guaranteed as we set $ N=32 $.

\section{Road Condition Monitoring} \label{RCM}
We will discuss rough road sensing and bump detection algorithms in this section. 
\subsection{Rough Road Sensing} \label{RR}
\begin{figure}[t]
	\centering
	\includegraphics[width=0.45\textwidth]{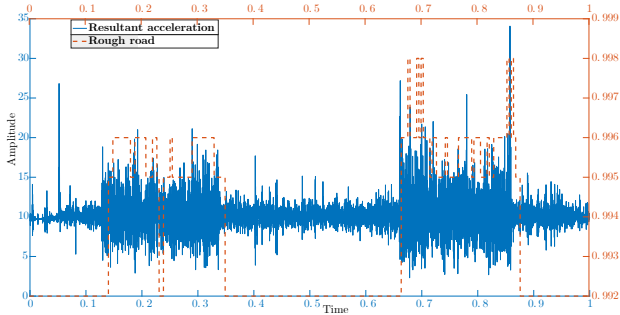}
	\caption{Sensed rough road and raw accelerometer signal.}	 
	\label{fig5}
	\vspace{-0.5cm}
\end{figure}
Cyclist feels discomfit and unsafe when rides on a rough road (\textit{i.e.} coarse or uneven road). Moreover, rough roads could induce false alarms of bumps unless it is adequately addressed. Therefore, the sensing of rough road is necessary for urban infrastructure developers. In this section, we will show that how the 3-axis accelerometer sensor on the smartphone is utilized to sense the rough road conditions. 

The relationship between the excessive vibration $  \sigma r $ generated by the rough road and $ a_{Rg} $ can be given as,
\begin{equation}\label{eqn13}
a_{Rg}[n]=\bar a_{Rg}[n] +  \sigma r[n],
\end{equation}
where $\bar a_{Rg}[n] $ is the undistorted accelerometer signal and $ \sigma r[n] \sim  N(\sigma,0) $ \textit{i.e.} we model the excessive vibration as additive Gaussian noise. Notice that the reason for existence of $ \sigma r[n] $ in Eqn. \ref{eqn13} is that cutoff frequency of the low pass filter, which is proposed in Section \ref{RRD} is insufficient to suppress the additive noise generated by rough roads. Therefore, the cutoff frequency should be adjusted (\textit{i.e.} decreased) when cyclist rides on a rough road and should be rebounded to its original value when rides on a smooth road. We accomplish this by estimating $\sigma $ by transforming $\hat a_{Rg}  $ into the wavelet domain and refining the cutoff frequencies accordingly.

The coefficients of the wavelet transform are usually sparse  \textit{i.e.} wavelet transformation of $ \hat a_{Rg} $ yields small coefficients that are dominated by noise and large coefficients, which carry more signal information. Moreover, the orthogonal wavelets have a decorrelation property \textit{s.t.} $ W\sigma r[n] = \sigma r[n]\sim  N(\sigma,0)  $ and since noise remains white in all bases, it does not influence the choice of basis. Therefore, we can accurately estimate the noise variance $ \hat \sigma $ from actual $ \sigma $ (which can be taken as the noise level of the signal) by transforming $ \hat a_{Rg} $  into wavelet domain.

When $ a_{Rg} $ is piecewise smooth, (which is the natural scenario) a robust  $ \hat \sigma $ is calculated from the median absolute deviation of the finest-scale wavelet coefficients is proposed in \cite{spatial} as follows,
\begin{equation}\label{eqn14}
\hat \sigma=\displaystyle \frac{\displaystyle\text{meadian}({|d_{j-1,k}|}:k = 0,1,.....,2^{j-1}-1)} {0.6745 },
\end{equation}
where the factor in the denominator is the scale factor which depends on the distribution of
$ d_{jk} $, and is equal to 0.6745 for a normally distributed data. In this paper, it is assumed that $\sigma $ is known, and for convenience it is taken to be 1. 
We propose a simple adaptive first order low pass \textit{IIR} filter, which adjusts its frequency response parameters based on the estimated noise variance $ \hat \sigma $ that can be calculated according to Eqn. \ref{eqn14}. It should be noted that although it is possible to estimate $\sigma $, a conventional adaptive \textit{IIR} filter is undesirable for our application as it is required the sign of the error signal (\textit{i.e. sign of} $|a_{Rg}[n]-\bar a_{Rg}[n]| $)\cite{IIR}. An adaptive filter approximate $ \bar a_{Rg} $ by minimizing $ \sigma r $ of observed $  a_{Rg} $ (\textit{i.e.} alternating its frequency response). As discussed in Section \ref{RRD}, frequency response of the filter is adjusted by altering the smoothing factor $ \alpha $. Therefore, we minimize $ \sigma r $ according to the following cost function $ J(\alpha) $.
\begin{equation}\label{eqn31}
J(\alpha) = \displaystyle \sum_{i=0}^{l}\lambda^{i-1} \hat \sigma r[i]
\end{equation}
where $0<\lambda\leq1$ is the forgetting factor and only $ l $ tapped delays of $\sigma r  $ are  considered to ensure a faster convergence. Moreover, we report the value of $ \alpha $ as an estimator for the roughness of the road  \textit{i.e.} a road is considered as rough whenever $ 0.992 < \alpha \leq 0.998 $. The adaptive scheme of the parameter $\alpha$ as follows,
\begin{equation}\label{eqn15}
\alpha = \begin{cases}
\enspace 0.998  & \text{if } J(\alpha) \in [0.01l,\infty) \\
\enspace 0.996  & \text{if } J(\alpha) \in [0.008l,0.01l) \\
\enspace 0.995  & \text{if } J(\alpha) \in [0.007l,0.008l) \\
\enspace 0.992  & \text{otherwise } 
\end{cases}.
\end{equation}
We classify road roughness levels based on the different values of $\alpha$. Therefore, the roughness of road is aggrandized as $\alpha$ increased. We understand that by adding more $\alpha$ levels will result in much smoother adaptiveness of the filter and more classification levels for rough roads. However, in this paper, we restrict $\alpha$ as we realize that it inessential for our application.

Fig. \ref{fig5} depicts the experiment results of our rough road sensing algorithm. We rode on a road that consists of two rough road segments and smooth road in between. Our algorithm identified the two rough road segments precisely, and the detected different rough road levels seem reasonable with our detailed inspection about the roughness of the road. Moreover, there could be a small difference between the actual starting point of the rough road and detected point. This could be due to the convergence rate of the algorithm and should be reasonable to neglect as it only accounts for few centimeters.
\subsection{Bump Detection} \label{BD}

\begin{figure*}[ht]
	\fontsize{8pt}{8pt}\selectfont
	\begin{tabular}{cc}		
		\includegraphics[width=0.45\textwidth]{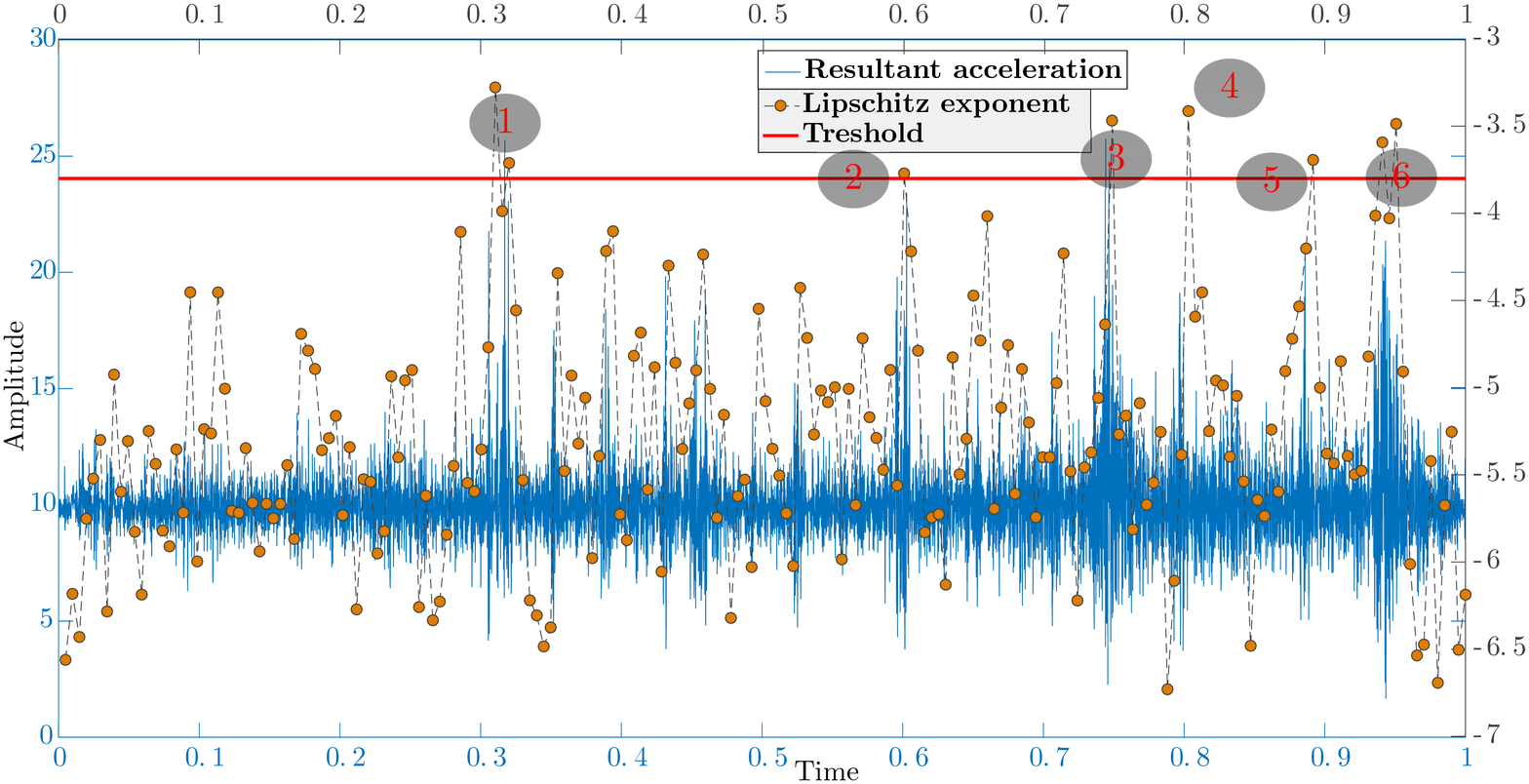}&   
		\includegraphics[width=0.45\textwidth]{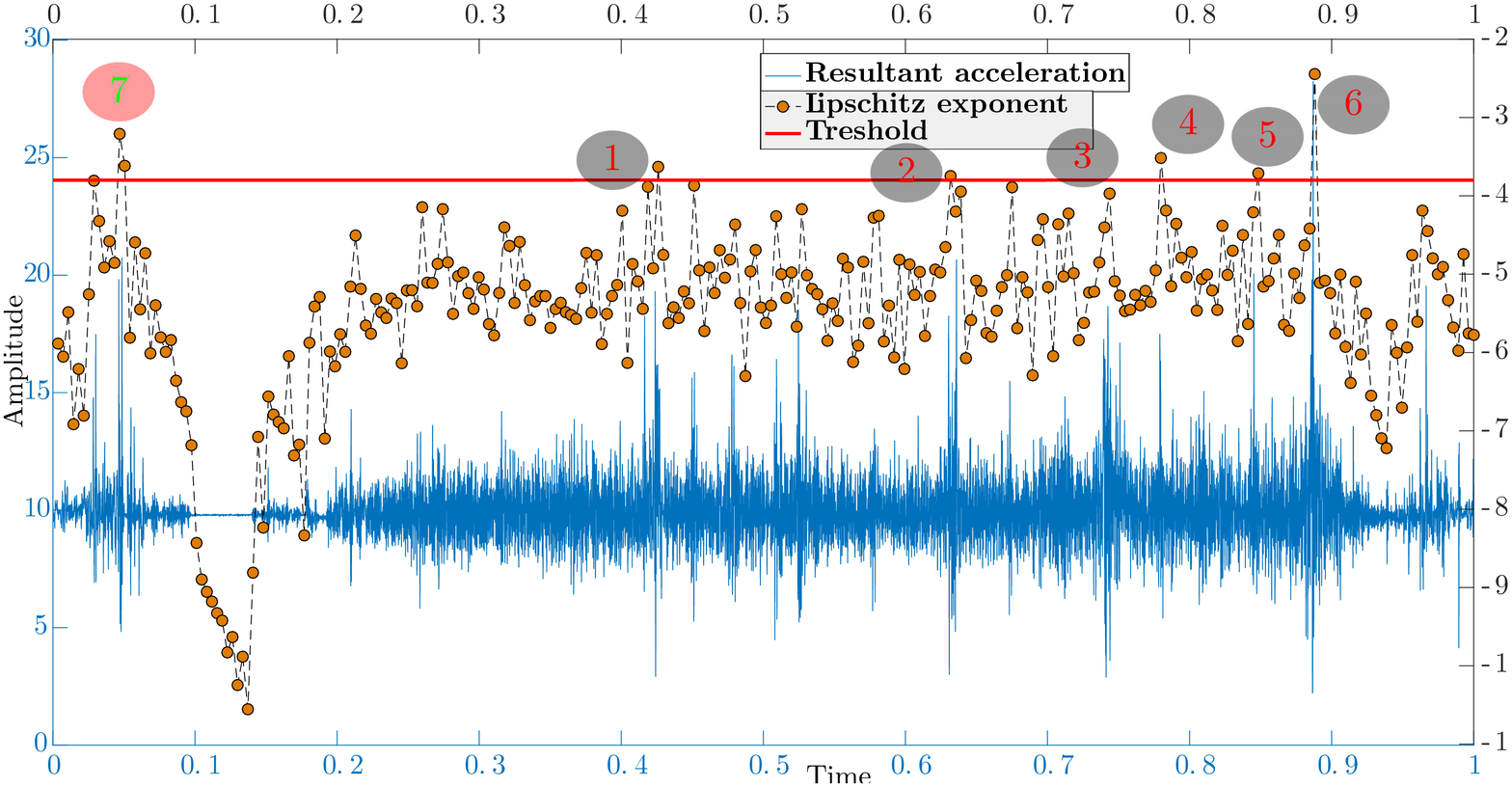}    
	\end{tabular}
	\caption{Detected bumps for \textit{Xiomi note 3} mounted (left) and  \textit{Samsung S3} placed in front bag (right).}
	\vspace{-0.5cm}
	\label{fig7}
\end{figure*}

As discussed in Section \ref{RW}, there are many different approaches for road bump detection in literature. Lack of suitability of these methods for our application are two-fold:
\begin{itemize}
	\item Most of these methods use fixed orientation of the accelerometer. Furthermore, as noted in Section \ref{ID}, there can be many different orientations of the accelerometer on smartphones that depend on the placements. Therefore, these placements sense the road bumps divergently.
	\item Different smart phones' accelerometers report discrepancy of their ranges. As shown  in Fig. \ref{fig7} (blue line), \textit{Xiomi Note 3} and \textit{LG 4G LTE} show significant divergence of their readings (for the same trip and same placement). 
\end{itemize}
By looking over above cases, we can notice that the method, which quantify the bump should not depend solely on the magnitude of the accelerometer readings. In this section, we will propose a bump detection algorithm where the above problems are being addressed. 

A bump could arise due to potholes, speed bumps, and patches, etc. Since the bump results in a significant surge in the accelerometer, any of aforementioned bumps shows a localized singularity (which means that it is not differentiable) behavior of $a_{Rg} $. Therefore, a bump can be detected by characterizing this singular structure. Moreover, to characterize the singular structure, it is necessary to precisely quantify the local regularity of $a_{Rg} $ \cite{wavelet}. \textit{Lipschitz exponents} provides regularity measurements (depends only on the singular structure not on the magnitude) at any time point $ v $  \textit{i.e.} if  $a_{Rg} $ has a singularity at $ v $, then the  \textit{Lipschitz exponent} at $ v $ characterizes the singular behavior. 

Suppose that $\hat a_{Rg} $ is $ m $ times differentiable in $ [v-h, v+h] $ and $a_{Rg}^{*}$ be the \textit{Taylor} approximation of $\hat a_{Rg} $ in the neighborhood of $ v $, then the \textit{Taylor} formula proves that the approximation error\cite{wavelet}
\begin{equation}\label{eqn16}
e_{v}(t)=\hat a_{Rg}(t) - a_{Rg}^{*}(t)
\end{equation}
that satisfies 
\begin{equation*}\label{eqn19}
\forall t \in  [v-h, v+h], \text{   }\text{   } |e_{v}(t)| \leq \displaystyle \frac{\displaystyle |t-v|^{m}} {\displaystyle m! }.
\end{equation*}
Hence, $ \hat a_{Rg} $ is said to be  \textit{Lipschitz} $ \beta \geq 0$ at $ t = v $ if there exists $ K > 0 $ and a polynomial $a_{Rg}^{*}$ of degree $ m $ ($ m $ is the largest integer satisfying $ m \leq \beta $) \textit{s.t.},
\begin{equation}\label{eqn17}
\forall t \in  \mathbb{R} \text{   } \text{   } \text{   }\text{   } |\hat a_{Rg}(t) - a_{Rg}^{*}(t)| \leq K|t-v|^{\beta}
\end{equation}
For instance, a function is not differentiable at $ t = v $ if $ 0 < \beta < 1 $. Then, the  \textit{Lipschitz exponent} $  \beta $ characterizes the nature of singularity at $ t = v $. Moreover, if we use a wavelet $ \psi(t) $ that has $ n>\beta $ vanishing moments \textit{i.e.},
\begin{equation}\label{eqn18}
\sum_{t} t^{k}\psi(t) = 0 \text{ }\text{ for }  \text{ } 0\leq k<n
\end{equation}
and by wavelet transformation applied, Eqn. \ref{eqn17} becomes,
\begin{equation}\label{eqn20}
W\hat a_{Rg}(j,k)=We_{v}(j,k)
\end{equation}
since $ Wa_{Rg}^{*}(j,k) = 0$, because of the vanishing moment property of the wavelet. \cite{smoothness} shows that, if $a_{Rg} \in L^{2}(\mathbb{R}) $ is Lipschitz $ \beta \leq n $ at $ v $, then there exists $ A $ \textit{s.t.},
\begin{equation}\label{eqn21}
\forall (j,k) \in \mathbb{R}\times \mathbb{R}^{+}, \text{   } \text{   } \text{   } |W\hat a_{Rg}(j,k)| \leq A j^{\beta +{1}/{2}} \displaystyle(1+|\displaystyle\frac{\displaystyle k-v}{\displaystyle j}|^{\beta})
\end{equation}
where, $|W\hat a_{Rg}(j,k)|  $ is the modulus maxima of $ \hat a_{Rg} $ at various scales $ j $. Near the cone of influence of $ t=v $ and by taking the logarithm, Eqn.\ref{eqn21} becomes \cite{detection},  
\begin{equation}\label{eqn22}
log_{2}|W\hat a_{Rg}(j,k)| \leq log_{2}A + (\beta + \frac{1}{2})log_{2}j
\end{equation}
Notice that by finding the slope of Eqn. \ref{eqn22} yields an estimate for $  \beta $. Although, simple linear regression simplifies the estimation procedure, it may be erroneous as it involves scaling the errors associated with the numerical estimation of the modulus maxima. Alternatively, a cost function $ J(A,\beta) $ can be derived from Eqn. \ref{eqn22} can be given as,
\begin{equation}\label{eqn23}
J(A,\beta) = \displaystyle \sum_{i,j=1}^{l} (log_{2}a_{i}-(log_{2}A+\beta log_{2}j))^{2}
\end{equation}
where $ a_{i}=|Wa_{Rg}(j,k)| $ for $ i,j=1,....,l$. Minimizing Eqn.\ref{eqn23} provides the following system \cite{lipschitz}
\begin{equation}\label{eqn24}
\resizebox{1\hsize}{!}{$%
\begin{pmatrix} A \\ \beta \end{pmatrix} = \begin{pmatrix} l &  \displaystyle \sum_{j=1}^{l} log_{2}j \\ \displaystyle \sum_{j=1}^{l} log_{2}j  & \displaystyle \sum_{j=1}^{l} (log_{2}j)^{2} \end{pmatrix}^{-1} \begin{pmatrix} \displaystyle \sum_{j=1}^{l} (log_{2}a_{i})\\\displaystyle \sum_{j=1}^{l} (log_{2}j)^{2} . \displaystyle \sum_{j=1}^{l} (log_{2}a_{i})\end{pmatrix}
$%
}%
\end{equation}
Notice that we only evaluate the $\beta$ from Eqn. \ref{eqn24}. Implementation of the  aforementioned procedure to evaluate $\beta$ could be tedious in Android. Therefore, we adopt a much simplified  algorithm for estimate $\beta$ (see Algorithm \ref{algo}), which provides a robust and pragmatic estimation (denoted as $\hat \beta$). Note that $\hat \beta$
is an estimate and hence is not necessarily equal to the true $\beta$.

\begin{algorithm}\label{algo}
	\SetKwInOut{Input}{Input}
	\SetKwInOut{Output}{Output}	
	
	\Input{$ \hat a_{Rg} $ }
	\Output{$\hat \beta$}
	$ L= 32 $                // Length of $\hat a_{Rg} $.\\
	$ levels = log_{2}(L) $  // Number of dyadic scales of $\hat a_{Rg} $.\\
	$M = \begin{pmatrix} 4&7 \\ 7 & 25 \end{pmatrix}^{-1}$ \\
	
	$ d_{1,k} = \text{Wavelet}(\hat a_{Rg}, levels ,1)$ // Wavelet transformation\\
	$ d_{2,k} = \text{Wavelet}(\hat a_{Rg}, levels ,2)$ \\
	$ d_{3,k} = \text{Wavelet}(\hat a_{Rg}, levels ,3)$ \\
	
	$[pks1,locs1] =\text{Findpeaks}(abs(d_{1,k}))$ // Peak value and its location index.\\ 
	$[pks2,locs2] =\text{Findpeaks}(abs(d_{2,k}))$ \\
	$[pks3,locs3] =\text{Findpeaks}(abs(d_{3,k}))$ \\

	\If{$\neg\text{isempty}(pks1)$}
	{
		$ [P1,I1] = \text{Max}(pks_1)$  // Maximum value and its location index. \\
		$location = locs1(I1)$\\
		$normloc1 = location/16  $\\
	}
	\If{$\neg\text{isempty}(pks2) \text{\bf AND} \neg\text{isempty}(pks1) $}
	{
		$ normloc2 = locs2./8 $\\
		$ normloc3 = normloc2-normloc1 $\\
		$ I2= \text{Min}(abs(normloc3)) $ // Minimum value\\
		$ P2 = pks2(I2); $  \\
	}
	
	$ \hat	\beta = M(2,1)*(log_{2}(P1)+ log_{2}(P2))+ M(2,2)*7*(log_{2}(P1)+ log_{2}(P2)) $\\
	
	return  $ \hat \beta $
	
	\caption{Algorithm for calculating \textit{Lipschitz exponent}}
	
\end{algorithm}


Fig. \ref{fig7} depicts the experimental results of our bump detection algorithm. We compared the results from \textit{Xiomi note 3} (\textit{SP1}) and \textit{Samsung S3} (\textit{SP2}), which were mounted and placed in a front bag, respectively throughout the trip. It is not hard to see that despite the different accelerometer readings between two smartphones, our algorithm successfully detected all 6 bumps in \textit{SP1} and 5 out of 6 bumps in \textit{SP2}. Note that we started recording the accelerometer readings from \textit{SP2} earlier and stopped later and hence the time stamps of same bumps should not be equal in Fig. \ref{fig7}. Moreover, the false bumps (number 7) that were detected in \textit{SP2} are due to the surge generated when placing the smartphone in the front bag. Although these type of false bumps are harder to avoid, we could remove that by comparing with \textit{GPS} velocity \textit{i.e.} by considering only the bumps that are detected when the cycle is moving faster than a certain threshold velocity. 

\section{CONCLUSION} \label{concl}
This paper studied an important smartphone application that passively monitors the road conditions of cycling routes. We described and evaluated the algorithms that sense rough cycling paths and locate road bumps. Moreover, we addressed multiple difficulties that limit such crowd-sourcing based smartphone application. We developed novel algorithms, which are based on wavelet transformation for rough road sensing and road bump detection. In rough road sensing, we estimated the excessive noise variance of the signal, which is generated by rough roads using wavelet transformation. Furthermore, we detect road bumps based on \textit{Lipschitz Exponent}, which is also calculated based on wavelet transformation. Our evaluations, which were done on cycling paths in Singapore has yielded promising results.

\bibliography{bibfile}

\begin{thebibliography}{10}

\bibitem{pothole}
J.~Eriksson, L.~Girod, B.~Hull, R.~Newton, S.~M.~R. Newton, S.~Madden, and
  H.~Balakrishnan, ``The pothole patrol: Using a mobile sensor network for road
  surface monitoring.,'' in {\em MobiSys '08 Proceedings of the 6th
  international conference on Mobile systems, applications, and services},
  June. 2008.

\bibitem{recursive}
M.~Ndoye, A.~M. Barker, J.~V. Krogmeier, and D.~M. Bullock, ``A recursive
  multiscale correlation-averaging algorithm for an automated distributed
  road-condition-monitoring system,'' {\em IEEE Transactions on Intelligent
  Transportation Systems}, vol.~12, pp.~795--808.

\bibitem{crowdsourcing}
J.~Chen, M.~Lu, G.~Tan, and J.~Wu, ``Crsm: Crowdsourcing based road surface
  monitoring,'' in {\em Proc. 11th IEEE/IFIP International Conference on
  Embedded and Ubiquitous Computing (EUC)}, Nov. 2013.

\bibitem{cond}
K.~Chen, M.~Lu, and X.~Fan, ``Road condition monitoring using on-board
  three-axis accelerometer and gps sensor,'' in {\em Proc. Communications and
  Networking in China (CHINACOM), 2011 6th International ICST Conference on},
  Aug. 2011.

\bibitem{time}
A.~Mednis, R.~Zviedris, G.~Kanonirs, and L.~Selavo, ``Real time pothole
  detection using android smartphones with accelerometers.,'' in {\em 2011 7th
  IEEE International Conference on Distributed Computing in Sensor Systems and
  Workshops}, June. 2011.

\bibitem{rich}
P.~Mohan, V.~N. Padmanabhan, and R.~Ramjee, ``Nericell: Rich monitoring of road
  and traffic conditions using mobile smartphones.,'' in {\em SenSys '08
  Proceedings of the 6th ACM conference on Embedded network sensor systems},
  November. 2008.

\bibitem{explore}
Douangphachanh and H.~Oneyama, ``Exploring the use of smartphone accelerometer
  and gyroscope to study on the estimation of road surface roughness
  condition,'' in {\em Proc. Informatics in Control, Automation and Robotics
  (ICINCO), 2014 11th International Conference on}, Sept. 2014.

\bibitem{speed}
V.~Rishiwal and H.~Khan, ``Automatic pothole and speed breaker detection using
  android system,'' in {\em Proc. 2016 39th International Convention on
  Information and Communication Technology, Electronics and Microelectronics
  (MIPRO)}, June. 2016.

\bibitem{ran_ieee_sensors2017}
R.~Liu, C.~Yuen, T.~N. Do, and U.-X. Tan, ``Fusing similarity-based sequence
  and dead reckoning for indoor positioning without training,'' {\em IEEE
  Sensors Journal}, vol.~17, pp.~4197--4207, July 2017.

\bibitem{Yassin_ieee_tutorials2016}
A.~Yassin, Y.~Nasser, M.~Awad, A.~Al-Dubai, R.~Liu, C.~Yuen, and R.~Raulefs,
  ``Recent advances in indoor localization: A survey on theoretical approaches
  and applications,'' {\em IEEE Communications Surveys Tutorials}, vol.~19,
  pp.~1327--1346, Secondquarter 2017.

\bibitem{classification}
M.~Hoffmann, M.~Mock, and M.~May, ``Road-quality classification and bump
  detection with bicycle-mounted smartphones.,'' in {\em UDM'13 Proceedings of
  the 3rd International Conference on Ubiquitous Data Mining}, Aug. 2011.

\bibitem{gravity}
D.~Mizell, ``Using gravity to estimate accelerometer orientatio.,'' in {\em
  Using gravity to estimate accelerometer orientatio}, November. 2003.

\bibitem{intro_wave}
C.~Leavey, M.~James, J.~Summerscales, and R.~Sutton, ``An introduction to
  wavelet transforms: a tutorial approach,'' {\em Insight \- Non-Destructive
  Testing and Condition Monitoring, J. Brit. Inst. Non-Destr. Test.}, vol.~45,
  no.~5, pp.~344--353, 2003.

\bibitem{waver}
G.P.Nason, {\em Wavelet Methods in Statistics with R}.
\newblock United Kingdom, UK: Springer, 2006.

\bibitem{nondy}
C.~Gupta, C.~Lakshminarayan, S.~Wang, and A.~Mehta, ``Non-dyadic haar wavelets
  for streaming and sensor data,'' in {\em Proc. IEEE 26th International
  Conference on Data Engineering}, pp.~569 -- 580, March. 2010.

\bibitem{spatial}
D.~L. Donoho and J.~M. Johnstone, ``Ideal spatial adaptation by wavelet
  shrinkage,'' {\em Biometrika}, vol.~81, no.~3, pp.~425 -- 455, 1994.

\bibitem{IIR}
J.~Shynk, ``Adaptive iir filtering,'' {\em IEEE ASSP Magazine}, vol.~6,
  pp.~4--21, April. 1989.

\bibitem{wavelet}
S.~Mallat, {\em A Wavelet Tour of Signal Processing}.
\newblock United Kingdom, UK: Elsevier, 2009.

\bibitem{smoothness}
S.~Jaffard, ``Pointwise smoothness,two-microlocalization and wavelet
  coefficients,''

\bibitem{detection}
J.-C. Hong, Y.~Kim, H.~Lee, and Y.~Lee, ``Damage detection using the lipschitz
  exponent estimated by the wavelet transform: applications to vibration modes
  of a beam,''

\bibitem{lipschitz}
H.~Izadi, K.~Innanen, and M.~P. Lamoureux, ``Continuous wavelet transforms and
  lipschitz exponents as a means for analysing seismic data,''

\end{thebibliography}
\end{document}